\documentclass[journal]{IEEEtran}
\ifCLASSINFOpdf
  \usepackage[pdftex]{graphicx}
  \graphicspath{{figures}}
\else
\usepackage[dvips]{graphicx}
  \graphicspath{{figures}}
\fi

\usepackage{amsmath}
\usepackage{amssymb}
 
 \ifCLASSOPTIONcompsoc
  \usepackage[caption=false,font=normalsize,labelfont=sf,textfont=sf]{subfig}
 \else
  \usepackage[caption=false,font=footnotesize]{subfig}
 \fi

\usepackage[acronym,toc,shortcuts]{glossaries}

\newacronym{QC}{QC}{quantum computing}
\newacronym{SQC}{SQC}{superconducting quantum computer}
\newacronym{SC}{SC}{superconducting}
\newacronym{HW}{HW}{hollow waveguide}
\newacronym{IR}{IR}{infrared}
\newacronym{FEM}{FEM}{finite element method}

\usepackage[
    backend=biber,
    style=ieee,
    sorting=ynt
]{biblatex} 
\begin{document}

\title{Low-pass filter with ultra-wide stopband for quantum computing applications}

\author{Robert Rehammar, Simone Gasparinetti
\thanks{Department of Microtechnology and Nanoscience, Chalmers University of Technology, Göteborg, Sweden}
\thanks{Coresponding author: Robert Rehammar, robert.rehammar@chalmers.se}
\thanks{Manuscript received Month xx, 2022; revised Month xx, 2022.}}

\maketitle

\glsresetall

\begin{abstract}
A new type of low-pass filter based on a leaky coaxial waveguide is presented. The filter has minimal insertion loss in the pass band, while at the same time high attenuation in the stop band is achieved. Thanks to its arrangement, the filter does not present parasitic leakage paths, so that, unlike conventional resonant filters, the stop band extends to very high frequencies. It is shown that a particular stop-band attenuation can be obtained by adding or removing leaking sections. Coupling between the center coaxial structure and the leaking holes is investigated. A prototype is manufactured and scattering parameters are measured up to 145 GHz. The prototype shows an insertion loss of less than 0.15 dB up to 10 GHz and an attenuation in excess of 60 dB above 70 GHz. The proposed filter is suitable for superconducting quantum computing applications where qubits are sensitive to radiation with energy high enough to break Cooper-pairs, thereby destroying superconductivity.
\end{abstract}


\glsresetall

\IEEEpeerreviewmaketitle


\glsresetall

\section{Introduction}
\label{sec:Introduction}
\Gls{SC} circuits using Josephson junctions to realize quantum bits (qubits) are a promising architecture for \gls{QC} \cite{devoret2013,bardin2020}. These devices are cooled to low temperatures (about 10 mK) to reduce thermal interference that can destroy superconductivity and interfere with qubit operations \cite{krinner2019,krantz2019}. Control electronics is placed at room-temperature and microwave signals are guided into successively cooler regions. At each temperature stage, the signal is attenuated to reduce thermally generated noise from higher temperature stages.

\begin{figure}[t]
	\centering
	\includegraphics[trim = 0mm 0mm 0mm 0mm, clip, width=0.95\columnwidth]{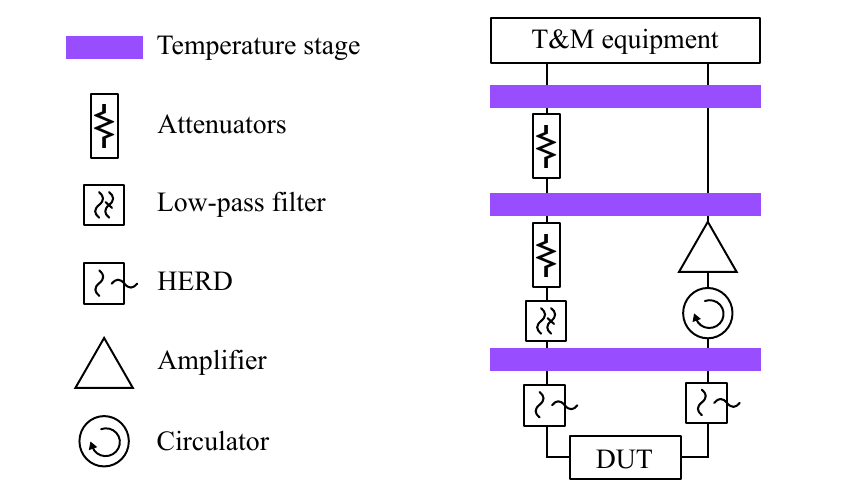}
	\caption{(color online) Simplified setup to operate a superconducting quantum processing unit, with test-and-measurement (T\&M) equipment at room temperature and a number of temperature stages reducing temperature to the base temperature where the device under test (DUT) resides, typically around 10 mK. Input lines are highly attenuated (for example, by 60 dB) in steps and output lines use circulators to block back-propagating radiation from higher stages. The High-Frequency Radiation Drain (HERD) filter presented in this work can be applied both at the input and at the output of the DUT.}
	\label{fig:architecture}
\end{figure}

\Gls{SC} qubits are peculiar in that they are not only sensitive to radiation in-band or radiation close to in-band frequencies (\gls{SC} qubits are commonly design to operate between 1 and 10 GHz) \cite{scigliuzzo2020}. They are also sensitive to very high-frequency radiation, even at extremely low power levels \cite{barends2011a,houzet2019a}. The reason for this is quantum mechanical: photons with energy high enough to break Cooper pairs can destroy superconductivity \cite{tinkham2004}. Cooper-pair breaking takes place when the frequency of the incident radiation exceeds $\frac{2 \Delta}{\hbar}$, where $\Delta$ is the superconducting energy gap and $\hbar$ is Planck's constant. Different materials and configurations have different energy gaps $\Delta$, but common materials and film thicknesses used in superconducting quantum computing have energy gaps corresponding to frequencies in the range of tens of GHz up to a few THz. For example, in Al, the most common material used to realize Josephson junctions, the threshold frequency for Cooper-pair breaking is around 80 GHz.
At the lowest temperature-stage, various filtering strategies have been implemented to remove any stray radiation, briefly reviewed below.
On output lines, circulators are used to block radiation from back-propagating from the high-temperature stages to the cold regions. A typical measurement setup is sketched in figure~\ref{fig:architecture}.

Designing filters that are transparent at low frequency (1 -- 10 GHz) and highly attenuating at very high frequencies ($\gtrsim 60$ GHz) is a non-trivial task. This type of filters are generally denoted \gls{IR} blocking filters in the \gls{QC} community. A simple LC-filter with discrete components has a transfer function
\[ f(\omega) = \frac{1}{1 - \omega^2 L C}. \]
However, the expression assumes ideal components. In reality, this type of filter will start leaking at some high-enough frequency due to parasitic elements in the components. Thus, other solutions have been utilized in \gls{QC} set-ups. A common approach is to use coaxial lines filled with some attenuating dielectric such as Eccosorb \cite{santavicca2008,slichter2009,danilin2022}, or copper-powder-filled epoxy \cite{martinis1987}. These filters are simple to manufacture and are highly attenuating at high frequencies. However, they have the drawback that they are also attenuating in-band frequencies. In addition, the attenuation is frequency dependent and increasing with increasing frequency. This can lead, among others, to undesired pulse distortion effects, which can cause errors when applying fast pulses to perform qubit gates \cite{simbierowicz2022}.

In this paper, an alternative design for a low-pass filter, based on a leaky coaxial waveguide, is proposed, which we named High-Energy Radiation Drain (HERD). HERD has excellent in-band performance, while at the same time, being highly attenuating at high frequencies.

The outline of this paper is as follows: Section II describes the design of the filter, Section III reports on measurement and simulation results, and Section IV concludes the paper.


\section{Design}
\begin{figure}%
\centering
\begin{subfloat}
	\centering
	\includegraphics[trim = 0mm 0mm 0mm 0mm, clip, width=0.95\columnwidth]{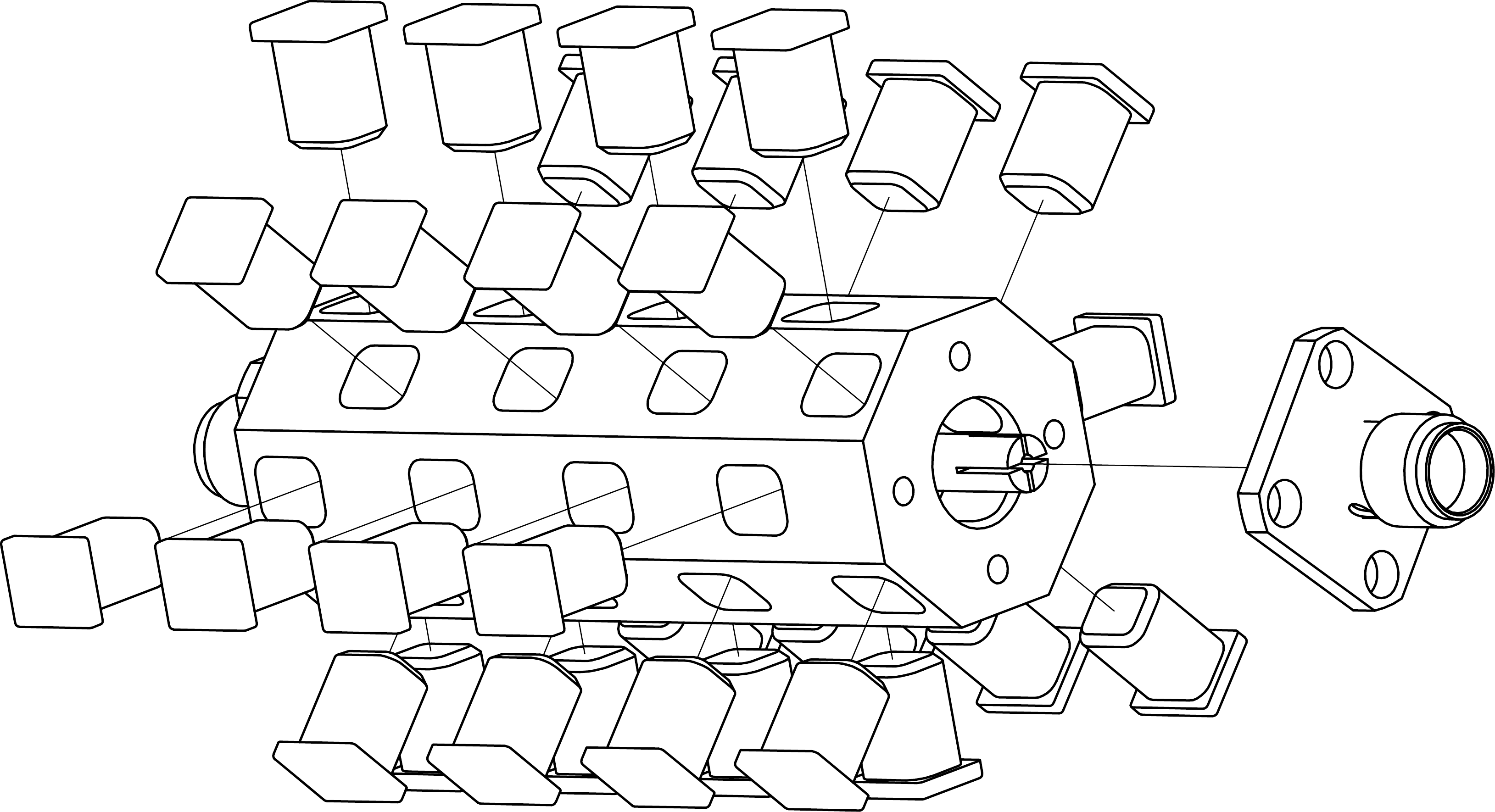}
	\label{fig:prototype_cad}
\end{subfloat}
\begin{subfloat}
    \centering
    \includegraphics[trim = 0mm 0mm 0mm 0mm, clip, width=0.95\columnwidth]{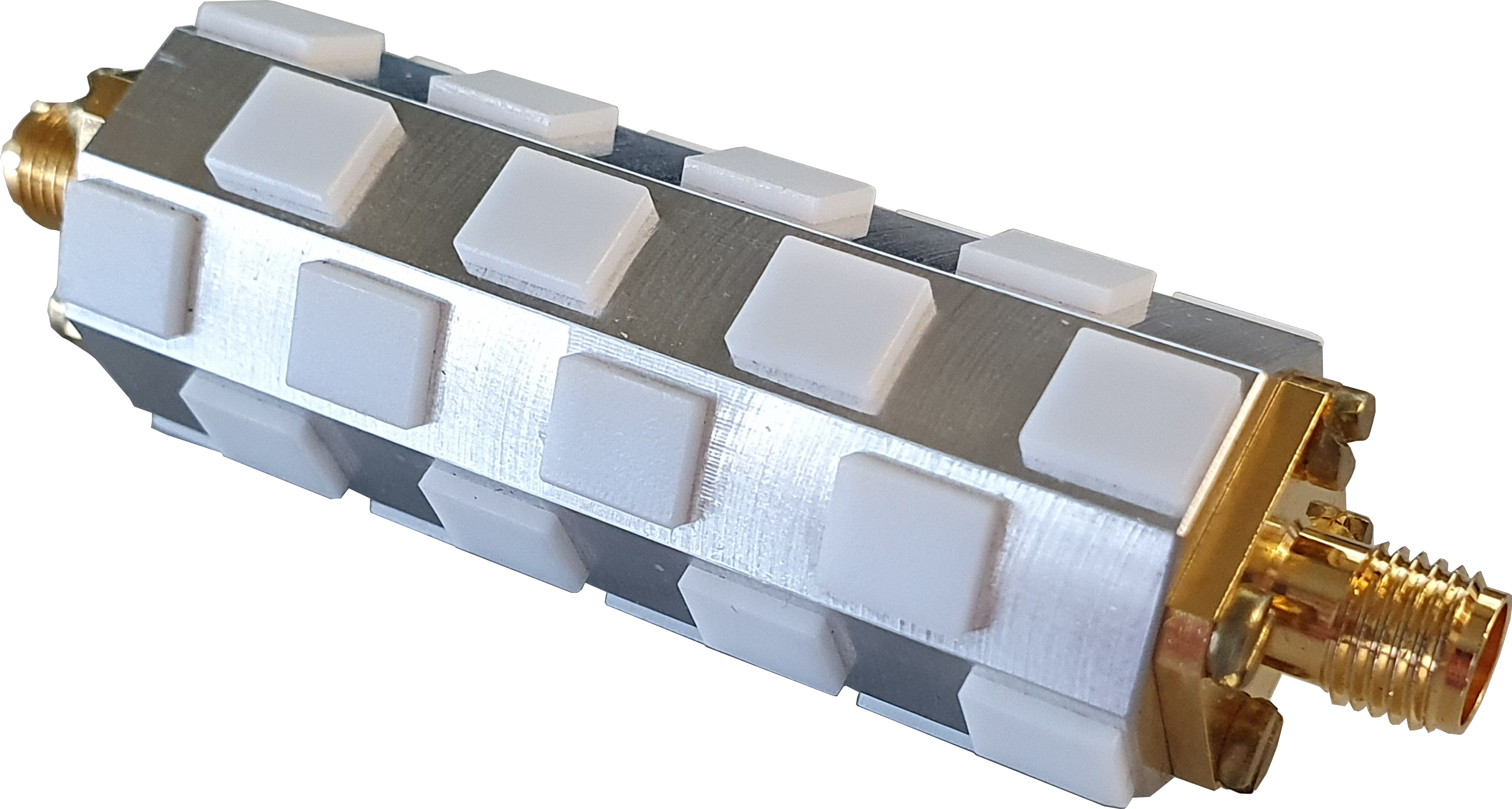}
	\label{fig:prototype_photo}
\end{subfloat}
\caption{(color online) (upper) Exploded view of the prototype. The knobs are the dielectric slabs that protrude into the waveguide. In the prototype, these are made from PTFE. (lower) Photo of the four-sections prototype.}
\label{fig:prototype_illustration}
\end{figure}

\begin{figure}[t]
	\centering
	\includegraphics[trim = 0mm 0mm 0mm 0mm, clip, width=0.95\columnwidth]{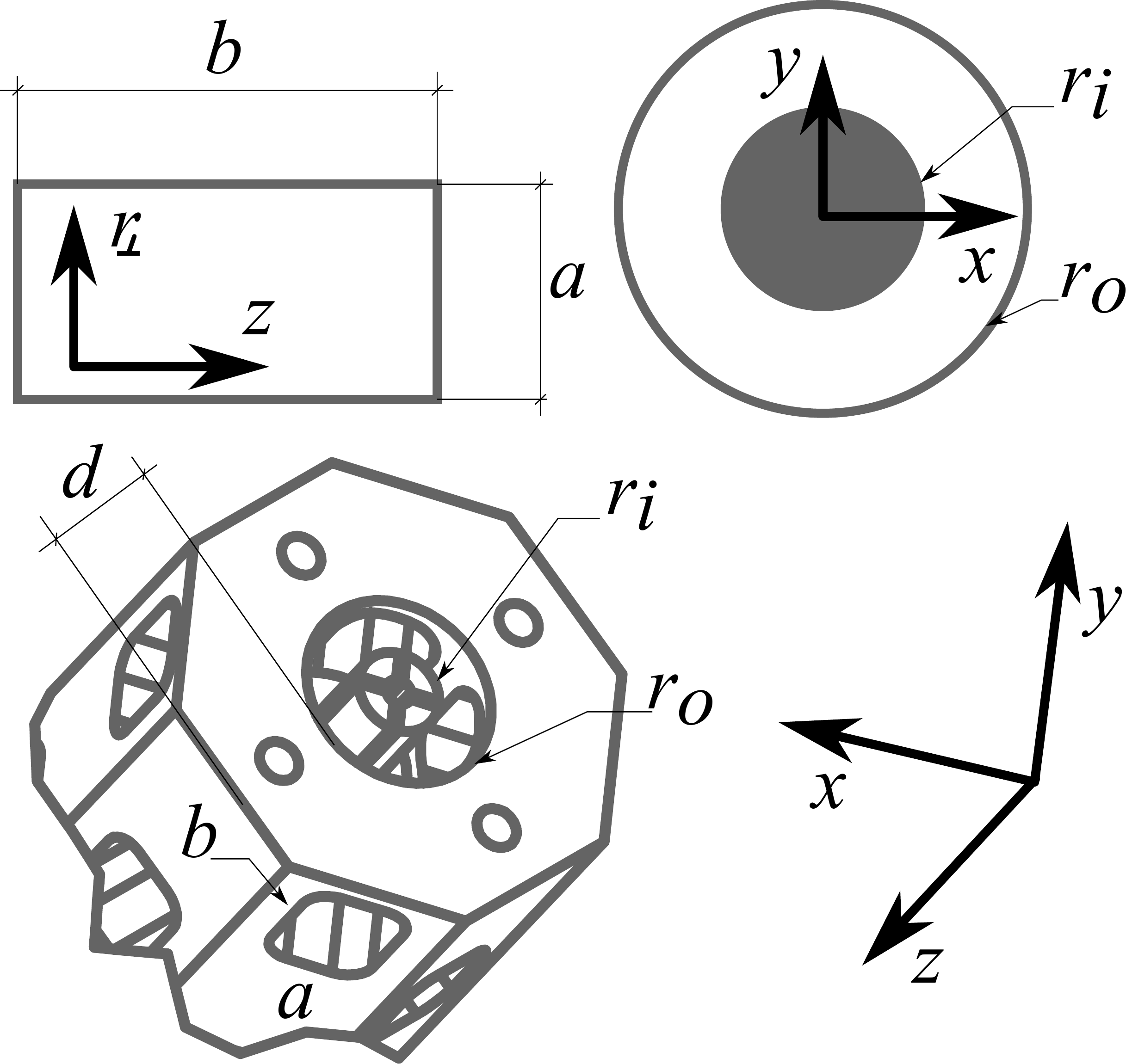}
	\caption{Dimensions and coordinate systems used in this paper of a rectangular waveguide (upper left), a coaxial transmission line (upper right) and the proposed filter (lower). The \gls{HW} dimensions are denoted {\em width} ($a$), {\em height} ($b$) and {\em depth} ($d$).}
	\label{fig:dimensions}
\end{figure}

HERD is built around a coaxial structure whose outer conductor presents waveguiding apertures to let radiation at high enough frequency leak out. Figure~\ref{fig:prototype_illustration} illustrates the prototype investigated in this paper and figure~\ref{fig:dimensions} shows parametric dimensions and coordinate systems used. A section is defined as a set of $2\times 4$ \gls{HW} apertures placed tangentially on the eight faces in two consecutive circles. $M$ is used to denote the number of sections and $M = 4$ in the prototype depicted in Figure~\ref{fig:prototype_illustration}. The total number of \gls{HW} is denoted $N$ and $N = 8 \times M$.
On the prototype and in all simulated design variants reported in this paper, the \gls{HW} are placed in a regular lattice on the coax outer body. This is purely for design and manufacturing simplification. To achieve the desired performance, the design of HERD does not rely on resonant phenomena or interference effects. Actually it is probably so that a non-periodic design might perform better, but that has not been investigated in this work.

The coaxial ports are the main waveguiding ports and they are terminated by SMA connectors. The characteristic impedance of a simple cylindrical coaxial transmission line is \cite{jackson1998}
\[ Z_0 = \frac{1}{2 \pi} \sqrt{\frac{\mu}{\varepsilon}} \ln\left(\frac{r_o}{r_i}\right) \simeq \frac{377~\Omega}{2 \pi} \ln\left(\frac{r_0}{r_i}\right) \]
where the last equality is obtained from assuming air as the dielectric. By setting $Z_0 = 50~\Omega$ with an air-filled dielectric
\begin{equation}
    \frac{r_o}{r_i} \sim  2.302\ .
    \label{eq:matching_radii}
\end{equation}

The cutoff frequency of the leaking apertures is set by the aperture cross-section and the dielectric constant of the material filling the cross-section. For a rectangular \gls{HW}, the cutoff wavenumber is \cite{collin1991}
\[ k_{c,mn}^2 = \frac{1}{\varepsilon \mu} \left[\left(\frac{m \pi}{a}\right)^2 + \left(\frac{n \pi}{b}\right)^2 \right] \]
and the propagation constant is
\[ \Gamma^2_{mn} = k_{c,mn}^2 - k_0^2 \ , \]
where $a$ and $b$ are waveguide cross section dimensions, $m,~n$ are integer mode numbers, and $k_0$ is the free-space wavenumber satisfying $k_0^2 = \omega^2 \varepsilon \mu$.

To achieve a particular radial width, $a$, of the leaking apertures, it is required to have a large enough outer radius, $r_o$, of the coaxial structure. At the same time, it is important to keep the coaxial line operating in single mode for in-band frequencies. A higher-order mode can start propagating when \cite{mooijweer1971}
\begin{equation}
    \lambda \simeq \pi (r_o + r_i)\ .
    \label{eq:multimode}
\end{equation}
Combining \eqref{eq:matching_radii} and \eqref{eq:multimode} and solving for $r_i$ for a cutoff frequency of $10$ GHz for the lowest higher-order mode gives $r_i \simeq 3~\mathrm{mm}$.

\subsection{Coupling between a coaxial transmission line and a hollow waveguide aperture}
To achieve high attenuation, strong coupling between the coaxial structure and the \gls{HW} apertures is wanted. Heuristically, from the aperture cross-section, it can be expected that larger \gls{HW} apertures couple stronger. To improve coupling, the \gls{HW} apertures can be made electrically larger by filling them with a dielectric with $\varepsilon > \varepsilon_0$. In this work, PTFE with $\varepsilon \simeq 2.2$ was used. The effect of filling the apertures can also be understood by considering that at higher frequencies, more radiation leaks through the apertures so that a larger insertion loss is observed. Increasing the electrical size of the \gls{HW} can thus be assumed to have a similar effect.

\begin{figure}[t]
\centering
    \includegraphics[trim = 0mm 0mm 0mm 0mm, clip, width=0.95\columnwidth]{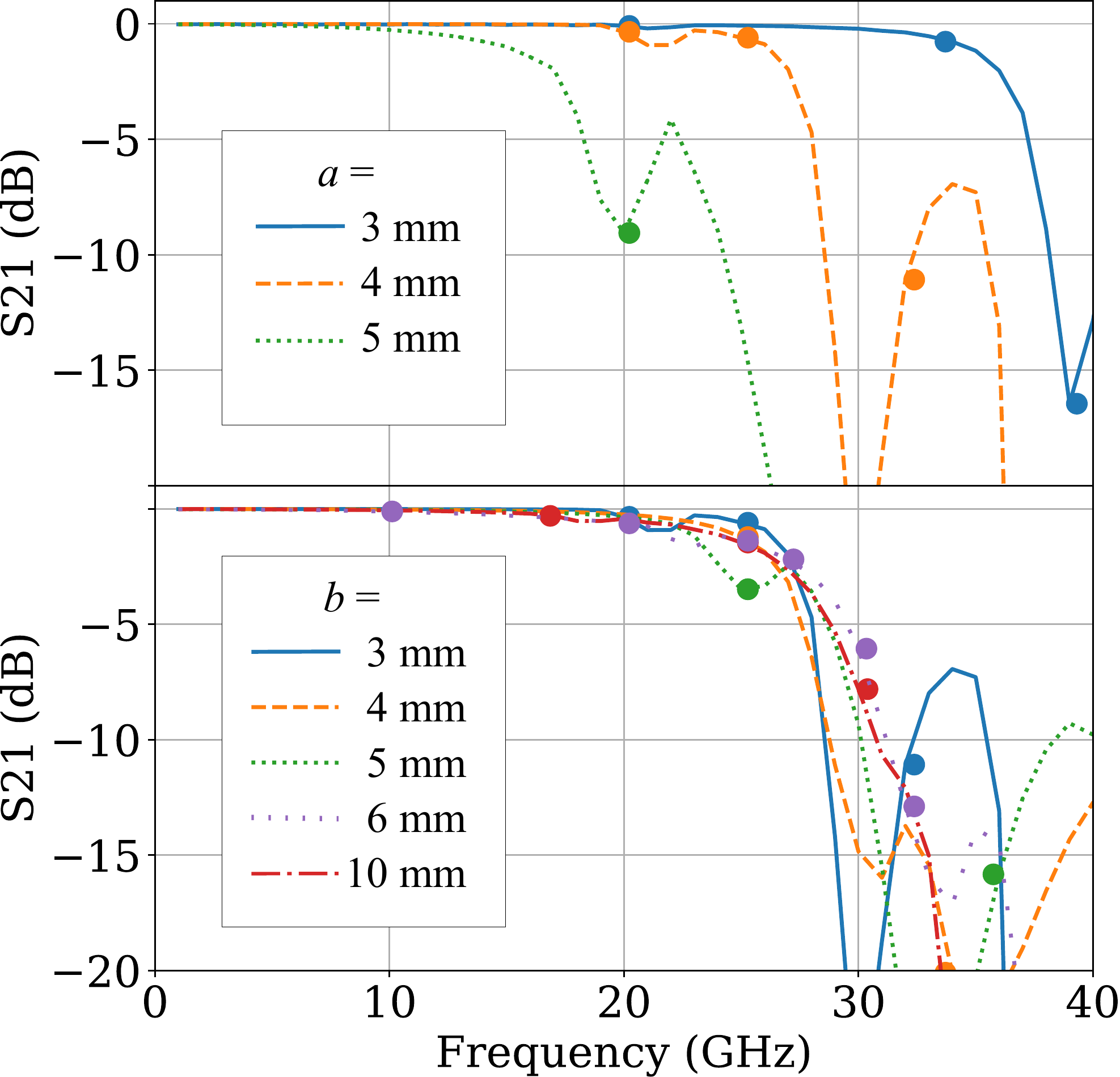}
    \caption{(color online) Simulated insertion loss of the filter when varying aperture width/height (top/bottom). Dots indicate the lowest order modes in the \gls{HW}s. Other parameters of the geometry are fixed according to Table~\ref{tab:prototype_parameters}.}
    \label{fig:coupling_dependence}
\end{figure}

The dependence of insertion loss on aperture dimensions $a$ and $b$ was investigated using \gls{FEM} simulations of the structure. The results are depicted in Figure~\ref{fig:coupling_dependence}. A very low, flat insertion loss is observed below a certain corner frequency. It can be seen that the corner frequency only depends on the width of the apertures, thus indicating that the excited mode in the \gls{HW} apertures is of TE$_{m0}$-type, at least in the lower frequency range.
At higher frequencies, both the coaxial and \gls{HW} structures are overmoded and can be expected to support complex superpositions of many modes. The irregular ripples seen in the insertion loss is expected to come from this fact. No effort is made to understand the exact nature of those modes and how they couple.

\subsection{Aperture depth}
Insertion loss can come from four main sources: mismatch (return loss), intrinsic metallic losses in the coaxial structure, dielectric losses in any exposed dielectric, and leakage through the \gls{HW}. It is interesting to investigate how these different loss mechanisms compare to find an optimum design.

Return loss of the prototype investigated in this work is around $-20$ dB, corresponding to about $0.04$ dB of insertion loss. This is a common design target of microwave components in general and a reasonable reference number.

Metallic losses will depend on choice of metal; in a cryogenic environment, they can be made negligible by constructing the filter housing in a material that becomes superconducting. As for dielectric losses, PTFE, used in the prototype as dielectric, has very low loss (loss tangent $< 0.0004$, \cite{thechemourscompanyfcllc.2018}) and is mostly not participating in transmission attenuation since it does not fill the coaxial structure.

\begin{figure}
    \centering
	\includegraphics[trim = 0mm 0mm 0mm 0mm, clip, width=\columnwidth]{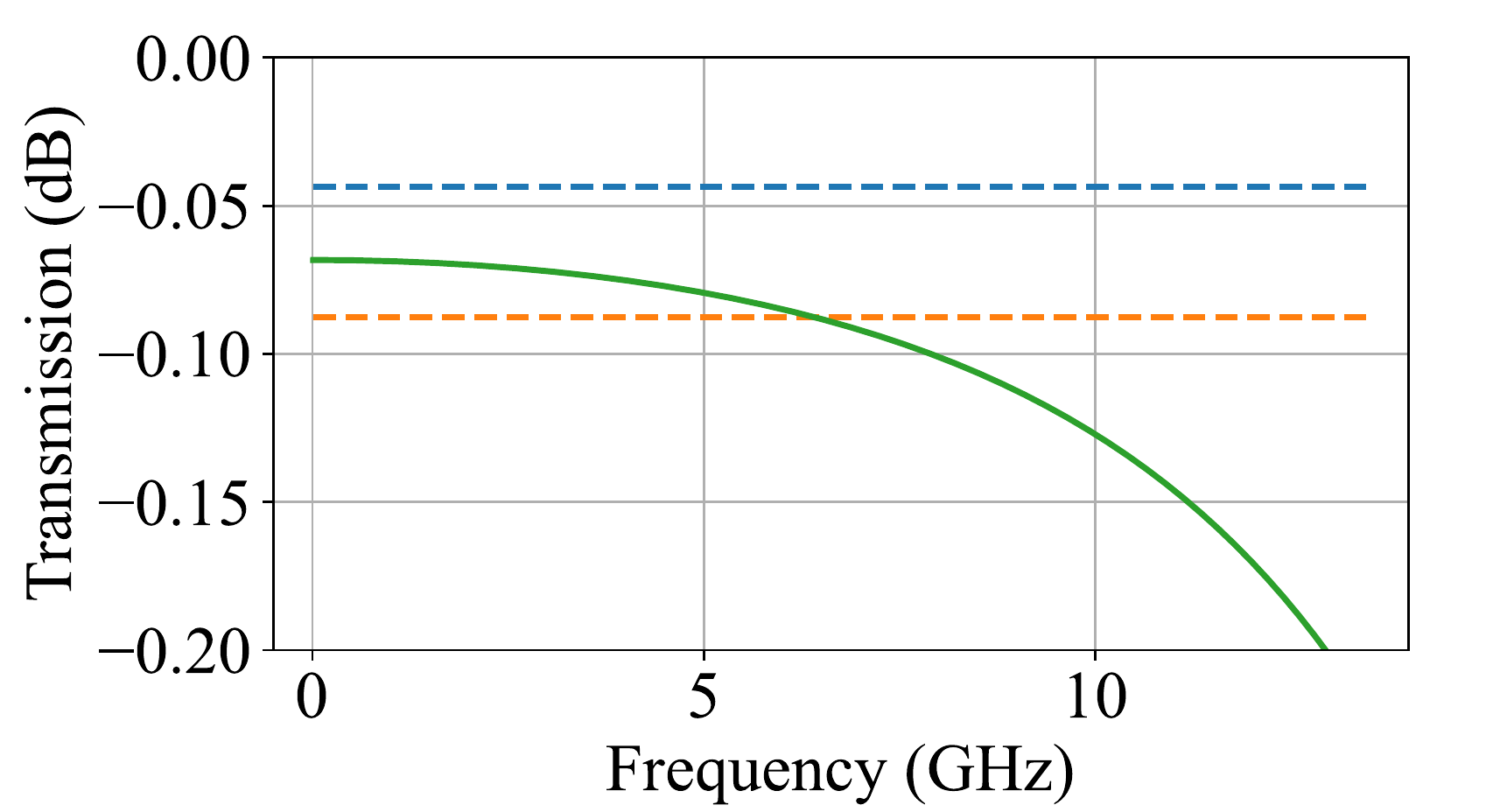}
    \caption{(color online) Computed in-band insertion loss (green/solid) from equation~\eqref{eq:leakage_loss} compared to mismatch loss (dashed) from $-23$ dB (blue) and $-20$ dB (orange).}
	\label{fig:computed_inband_loss}
\end{figure}

The aperture \gls{HW} needs to have a sufficient depth to stop low-frequency (in-band) radiation from tunneling through the apertures. Below cutoff, the field amplitude is attenuated according to
\[ F(d) = e^{-\gamma d} \]
where
\[ \gamma = \sqrt{\left( \frac{m \pi}{a} \right)^2 + \left( \frac{n\pi}{b} \right)^2 - \omega^2/c^2} \]
where $m,n$ is the mode number, $a,~b$ are aperture dimensions as defined in Figure~\ref{fig:dimensions} and $c$ speed of light in the medium. The lowest-order mode is attenuated the most and it is assumed that the TE $m,~n = 0,~1$ mode couples strongest to the TEM mode of the center waveguide as indicated by the above analysis. A simple loss model where loss is additive in the number of apertures is
\begin{equation}
    T = \left(1 - |F|^2\right)^M
    \label{eq:leakage_loss}
\end{equation}
where $M$ is the number of apertures (Note, $M = 8 \times N$). The transmission, $T$, for a filter with parameters according to table~\ref{tab:prototype_parameters} is plotted in Figure~\ref{fig:computed_inband_loss}.
It can be seen that in-band insertion loss can be dominated by leakage through the \gls{HW} apertures, depending on their parameters. There is a trade-off between HERD size and \gls{HW} depth that is application dependent since a larger aperture depth, $d$ improves performance at the expense of filter size. This analysis shows that the aperture geometry needs to be engineered properly to achieve desired performance.

\section{Results}
In this Section, results from investigations of a complete HERD filter are presented. Parameter values used in the simulations and fabricated prototype can be found in table~\ref{tab:prototype_parameters}.
\begin{table}[ht]
    \centering
    \begin{tabular}{|rrl|}
        \hline
        aperture width & $a$ & 4 mm \\
        aperture height & $b$ & 5 mm \\
        aperture depth & $d$ & 4.85 mm \\
        coax outer radius & $r_o$ & 3.65 mm\\
        coax inner radius & $r_i$ & 1.59 mm\\
        slab dielectric constant & $\varepsilon_{\mathrm{slab}}$ & 2.2\\
        number of sections & $N$ & 4\\
        \hline
    \end{tabular}
    \caption{Prototype parameters.}
    \label{tab:prototype_parameters}
\end{table}

\subsection{Simulation}
The filter prototype design is investigated in detail using frequency domain \gls{FEM} simulations as implemented in COMSOL Multiphysics$\textsuperscript{\textregistered}$ \cite{comsol6.2022}. The metal is modeled as a perfect electric conductor. In the simulation model, the apertures are terminated by scattering boundary conditions to make the simulations more efficient. In the manufactures prototype, the \gls{HW}'s are terminated by an absorbing foam or open air. The exact details of the termination have small effects on the filter performance as long as it is sufficiently attenuating.
\begin{figure}[t]
    \centering
    \includegraphics[trim = 0mm 0mm 0mm 0mm, clip, width=0.95\columnwidth]{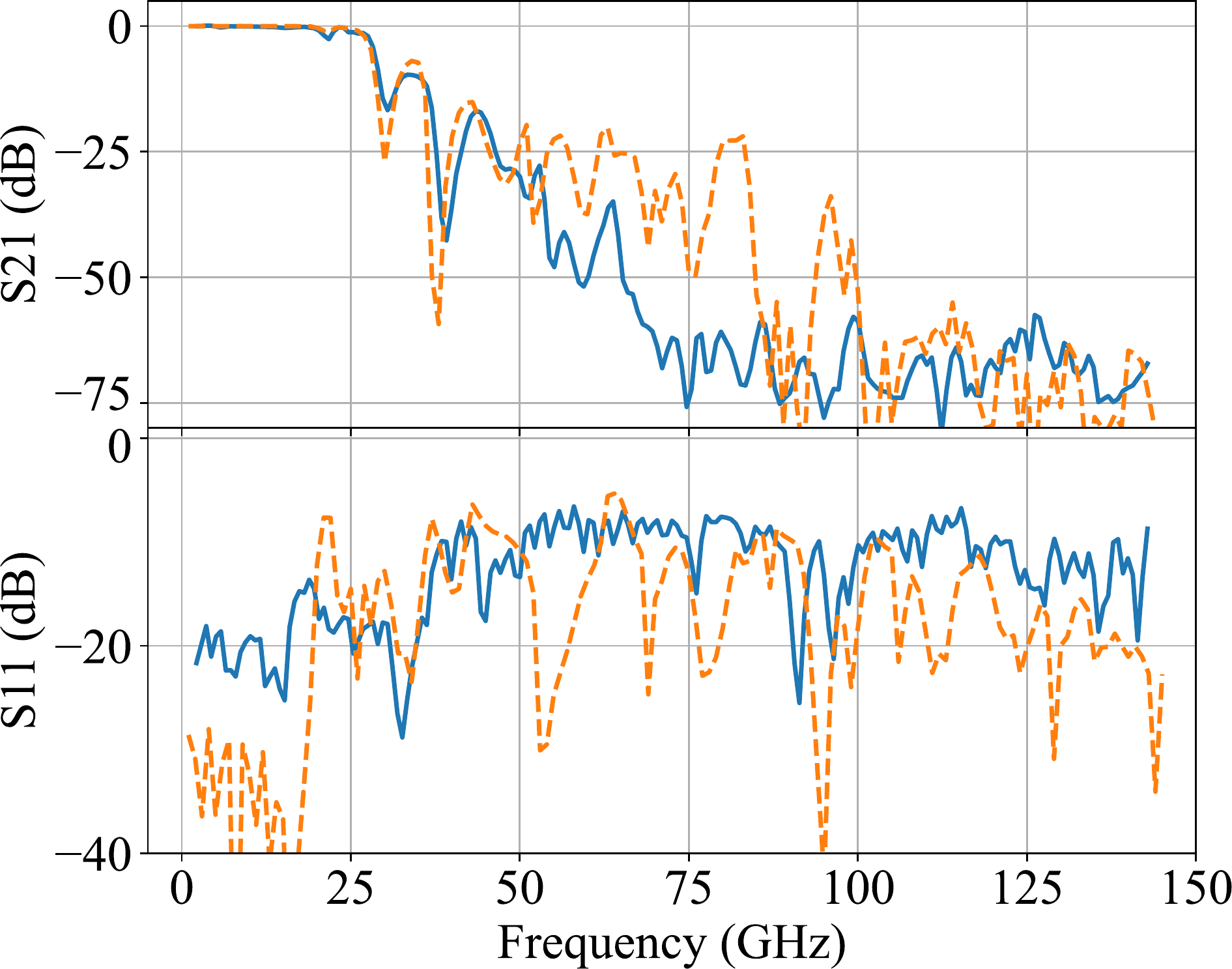}
    \caption{(color online) Measured (solid blue) and simulated (dashed orange) high frequency scattering parameters (top) $S_{21}$ (bottom) $S_{11}$ of the filter.}
    \label{fig:full_band}
\end{figure}
The results can be found in Figure~\ref{fig:full_band} (solid lines).
Around 80 and 100 GHz in the simulation model the attenuation is much reduced. These features are not present in the measured prototype. They are ascribed to the scattering boundary conditions of the \gls{HW} in the simulation.

\begin{figure}[t]
    \centering
	\includegraphics[trim = 0mm 0mm 0mm 0mm, clip, width=\columnwidth]{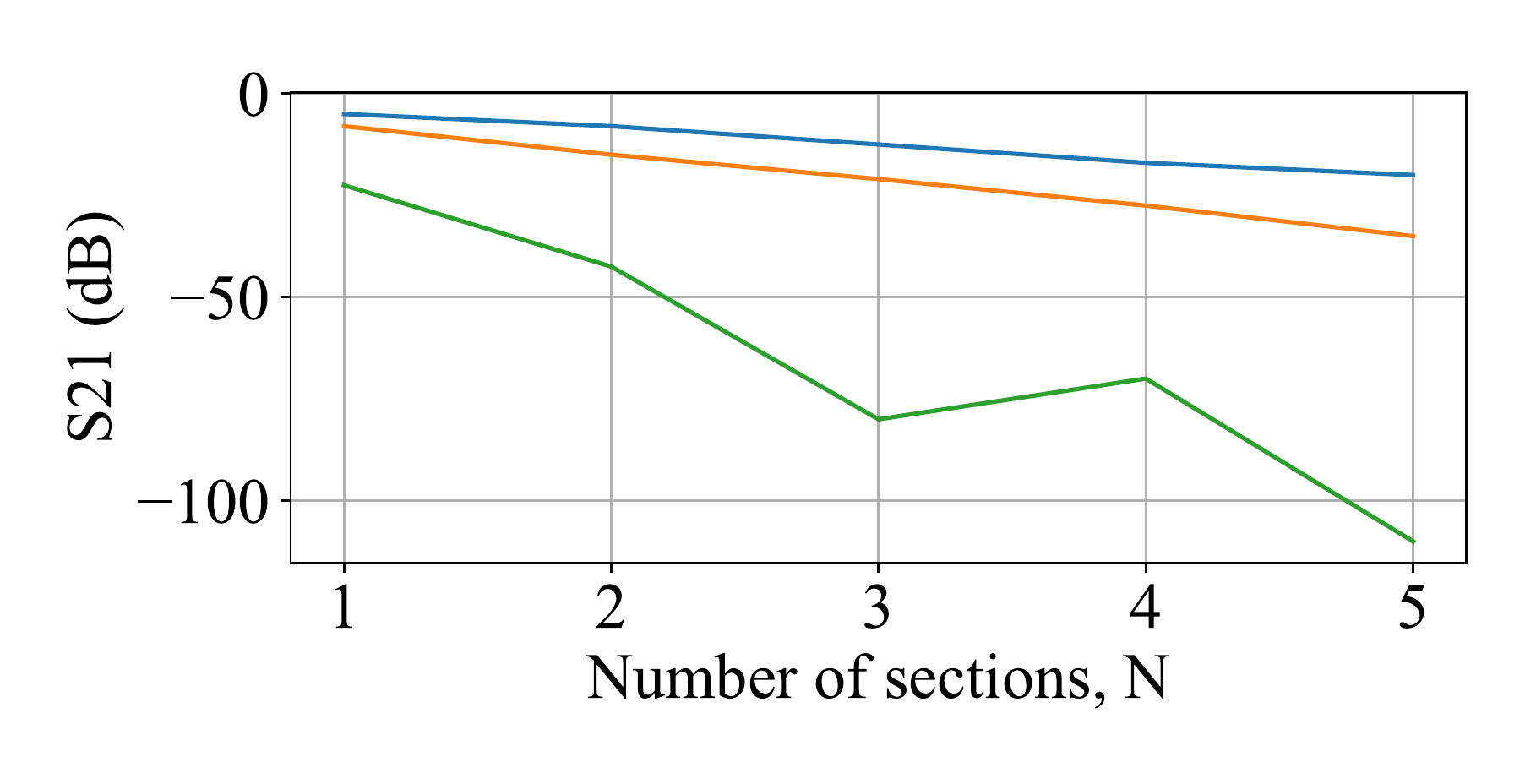}
    \caption{(color online) Attenuation scaling from \gls{FEM} simulations with number of sections for a set of frequencies, 40 GHz (blue), 60 GHz (orange) and 130 GHz (green).}
    \label{fig:attenuation_from_sections}
\end{figure}

Figure~\ref{fig:attenuation_from_sections} (solid lines) shows simulation results on how attenuation through the filter scales with the number of sections for a set of frequencies. It can be seen that a particular attenuation can be achieved by adding sections until sufficient attenuation is achieved. For high frequencies, the attenuation is not monotonic in the number of sections. That is an indication of the complex electromagnetic environment at such high frequencies.

\subsection{Measurements}
The prototype, depicted in figure~\ref{fig:prototype_cad}, was measured using a 145 GHz VNA. The filter structure is highly overmoded in large parts of the band of interest. To measure transmission through the filter, which is equipped with SMA connectors, several adapters were used to convert the 0.8 mm coaxial ports of the VNA to 2.9 mm ports compatible with the SMA connectors on the filter prototype. A basic through calibration was performed using a 2.9-2.9 mm adapter connected to the 2.9 mm ports. The primary interest for this measurement is in the high-frequency attenuation of the filter. From the basic calibration procedure, the phase of the scattering parameters cannot be measured, but its measurement is inessential to the characterization of the filter performance in the stop band anyway.

Measurement and simulation results over the full 145 GHz span are depicted in figure ~\ref{fig:full_band}. A very good agreement is found between simulation and measurement except for the return loss, which is lower in the prototype. This is ascribed to the connectors which are not included in the simulation.
\begin{figure}[t]
	\centering
    \includegraphics[trim = 0mm 0mm 0mm 0mm, clip, width=0.95\columnwidth]{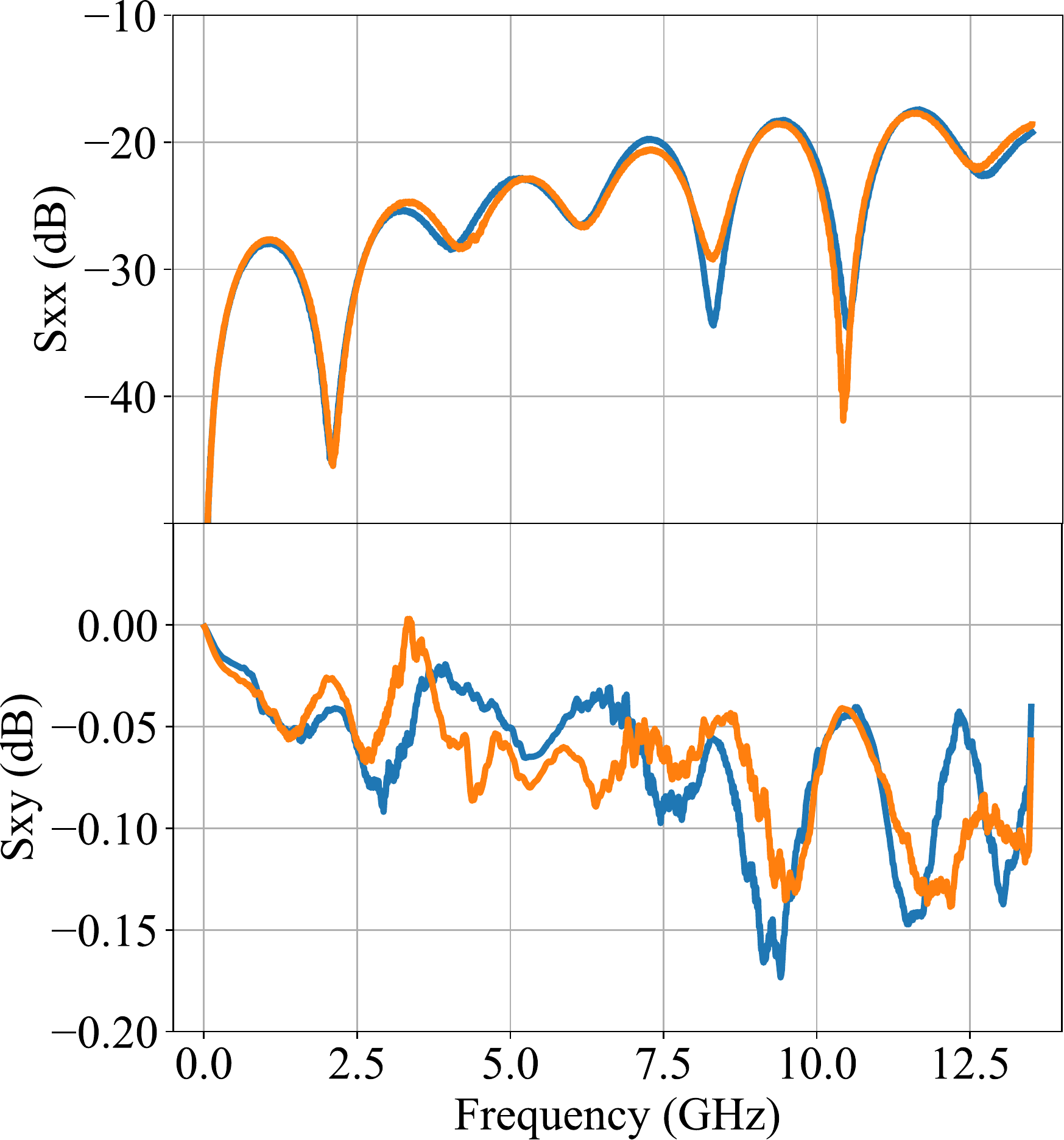}
    \caption{(color online) Measured in-band performance of the filter. (Top) Reflection coefficients, $S_{11}$ (blue) and $S_{22}$ (orange). (Bottom) Transmission coefficients, $S_{21}$ (blue) and $S_{12}$ (orange).}
    \label{fig:inband}
\end{figure}

The in-band performance of the filter is also of interest and was measured using a conventional VNA procedure and calibration. The results from the measurement can be found in figure~\ref{fig:inband}. The measured insertion loss is below 0.15dB up to 12 GHz and the ripples are well below 0.1dB in the 4-8 GHz band typically used for SC qubits.

\section{Conclusion and outlook}
In this paper, a new type of low-pass filter, named HERD, was presented in which coupling between a coaxial transmission line and apertures in the outer conductor is used to achieve attenuation above cutoff of the apertures. It was shown that a filter consisting of 4 sections and an outer radius $r_o = 3.65$ mm of the coaxial transmission line can provide over 60 dB of attenuation above 70 GHz while at the same time providing very small insertion loss below 10 GHz. Measurement results of a prototype filter up to 145 GHz were presented, in good agreement with \gls{FEM} simulations. To the authors knowledge, it is the first time such high-frequency measurements are presented on \gls{IR} blocking filters for \gls{QC} applications. It is worth noting that the filter stays reasonably matched all that way to 145 GHz, despite that the coaxial structure is highly overmoded at these frequencies.

The presented design can be further optimized and tailored to specific applications. Attenuation can be increased or decreased by increasing or decreasing the number of leaking sections. Several design parameters, such as number or sections and aperture shape, can be modified to improve performance. One example of a design improvement that was not investigated in this work is to shape the apertures as ridged waveguides. That can be expected to increase the bandwidth of the aperture and at the same time increase cutoff of some higher-order modes. In the prototype investigated in this work, \gls{HW} apertures were arranged  in a regular lattice along the coax transmission line. Possible resonances due to this periodicity were not investigated and might have degraded performance at specific frequencies. If required, excellent isolation and a steeper roll-off at frequencies intermediate between the passband and the stopband can be straightforwardly achieved by combining the presented filter with commercial, cascaded LC filters.

The authors expect HERD to find wide use in cryogenic setups for quantum computing applications, where high rejection of infrared radiation in the 100 GHz range is required. The extremely low insertion loss in the passband makes the filter amenable to be used as an isolation element between the devices under test and the first amplifier in the qubit readout chain. The very low ripples in the passband (less than 0.15dB over 10 GHz) make it also a competitive choice for all types qubit drive lines, in which absorptive filters are typically employed, especially when high-bandwidth, low-distortion pulses are required to execute qubit gates with high fidelity.

The filter is patent pending.


\section*{Acknowledgment}
The authors like to acknowledge support from Knut and Alice Wallenberg foundation via the WACQT program. We would also like to thanks to Lars Jönsson for manufacturing of the prototype.

\printbibliography[
heading=bibintoc,
title={References}
]




\end{document}